\documentclass[journal,12pt,onecolumn,draftclsnofoot,]{IEEEtran}
\usepackage{amsfonts}
\usepackage{amsbsy}
\usepackage{amsmath,amssymb}
\usepackage{amsthm}
\usepackage{times}
\usepackage{graphicx}
\usepackage{enumerate}
\usepackage[usenames]{color}
\usepackage[dvips]{pstcol}
\usepackage{epstopdf}
\usepackage{cite}
\usepackage{amsmath}
\usepackage{amssymb}
\usepackage{amsfonts}
\usepackage{graphicx}
\usepackage{epsfig}
\usepackage{psfrag}
\usepackage{xcolor}
\usepackage{amsfonts, bm}
\usepackage{epstopdf}
\usepackage{cite}
\usepackage{color}
\usepackage{cuted}
\usepackage{xcolor}
\usepackage{verbatim}
\usepackage{algorithm}
\usepackage{algorithmic}
\usepackage{booktabs}
\usepackage{colortbl}
\usepackage{caption}
\usepackage{subcaption}

\input epsf
\DeclareMathOperator*{\argmax}{arg\,max}

\begin{document}

\title{Distributed-Training-and-Execution Multi-Agent Reinforcement Learning for Power Control in HetNet
}
\author{Kaidi Xu, Nguyen Van Huynh, and Geoffrey Ye Li

\thanks{
The authors are with the Department of Electrical and Electronic Engineering, Imperial College London, London SW7 2AZ, U.K. (e-mail: k.xu21@imperial.ac.uk; huynh.nguyen@imperial.ac.uk; geoffrey.li@imperial.ac.uk)}}

\maketitle

\begin{abstract} 

In heterogeneous networks (HetNets), the overlap of small cells and the macro cell causes severe cross-tier interference. 
Although there exist some approaches to address this problem, they usually require global channel state information, which is hard to obtain in practice, and get the sub-optimal power allocation policy with high computational complexity. 
To overcome these limitations, we propose a multi-agent deep reinforcement learning (MADRL) based power control scheme for the HetNet, where each access point makes power control decisions independently based on local information. 
To promote cooperation among agents, we develop a penalty-based Q learning (PQL) algorithm for MADRL systems. 
By introducing regularization terms in the loss function, each agent tends to choose an experienced action with high reward when revisiting a state, and thus the policy updating speed slows down. 
In this way, an agent's policy can be learned by other agents more easily, resulting in a more efficient collaboration process. 
We then implement the proposed PQL in the considered HetNet and compare it with other distributed-training-and-execution (DTE) algorithms. 
Simulation results show that our proposed PQL can learn the desired power control policy from a dynamic environment where the locations of users change episodically and outperform existing DTE MADRL algorithms.

\end{abstract}
\vspace{-1em}

\section{Introduction}

Modern wireless networks require an enormous amount of data traffic to support a huge number of wireless devices with data-hungry applications like augmented reality (AR), virtual reality (VR), machine learning services, and ultra-high definition video streaming \cite{agiwal2016next, gupta2015survey, zhang2019visions}.
In conventional cellular networks, a macro base station (BS) needs to provide access to the core network for all user devices (UDs) in the cell.
To offload the heavy wireless traffic from those macro BSs, heterogeneous network (HetNet) is a promising technique \cite{xu2021survey}. 
In HetNets, different-tier access points (APs) coexist in one macro cell, where a macro AP is deployed in the center of the cell.
Small-cell APs, including femto-cell and pico-cell APs, are deployed in the inner regions to offload data traffic or at the macro cell edges to enhance the coverage.
These densely-deployed low-power and low-cost small-cell APs can increase the spatial and the spectrum efficiency but suffer from severe cross-tier and co-tier interference.
As a result, interference management has attracted great research interests.

\subsection{Current Approaches and Limitations}
To suppress interference, various approaches \cite{luo2008dynamic,shi2011iteratively, shen2018fractional, yang2016advanced,singh2013joint} have been developed in the literature to address the power control problem, which usually formulate the power control problem as a non-convex and NP-hard optimization problem \cite{luo2008dynamic} and then solve it by iterative algorithms \cite{shi2011iteratively, shen2018fractional}.
These centralized algorithms can find a sub-optimal solution but face the high computational complexity and scalability problems.
In addition, these centralized optimization-based algorithms require global channel state information (CSI) in advance, which is usually not available in multi-cell systems.
As a result, these conventional algorithms are not suitable for HetNets with real-time power control constraints.
This motivates significant research interests in deep learning (DL) based low-complexity and distributed power control solutions.

Several DL-based power control solutions have been proposed in \cite{sun2018learning, liang2019towards,hu2020iterative,chowdhury2021unfolding}.
By exploiting supervised learning, the neural network (NN) in \cite{sun2018learning} learns the mapping from available information to desired optimal or sub-optimal solutions, which are generated by traditional optimization algorithms.
Similarly, the power control neural network (NN) in \cite{liang2019towards} directly maximizes the system sum-throughput in a data-driven unsupervised learning manner.
Differently, the model-driven NN in \cite{chowdhury2021unfolding} efficiently allocates transmit power in cellular networks.
These DL-based power control algorithms can successfully reduce the complexity since they only need to forward propagate through the NNs to predict the transmit power.

Even using unsupervised learning, the above approaches require interference link CSI, especially inter-cell interference CSI, which is difficult to estimate in real systems.
To address the issue, various deep reinforcement learning (DRL) based methods \cite{nasir2019multi,liang2019spectrum,zhang2020deep,nasir2021deep,meng2019power,ye2019deep} have been proposed for power control in multi-cell systems. 
For example, the policy NNs in \cite{nasir2019multi} and \cite{nasir2021deep} are trained with shared trainable parameters for homogeneous wireless networks.
In particular, by exploiting information exchange among neighboring APs and UDs, a multi-agent DRL (MADRL) based power control algorithm has been developed in \cite{nasir2019multi} for homogeneous cellular networks, where a training center has been employed to train the common NN parameters for all agents,
while the deep deterministic policy gradient algorithm \cite{lillicrap2015continuous} jointly allocates power and sub-spectrum for cellular networks in \cite{nasir2021deep}.
To address the scalability issue, the MADRL-based power control scheme in \cite{liang2019spectrum} allocates transmit power and sub-channels in a distributed manner in a vehicle-to-vehicle network, where fingerprint \cite{foerster2017stabilising} is employed to stabilize the multi-agent system.
The  multiple-actor-shared-critic (MASC) method in \cite{zhang2020deep} trains the power control policy NNs deployed at each AP in HetNet.

Although achieving good performance, all these aforementioned DRL-based resource allocation algorithms are centralized-training-and-decentralized-execution (CTDE) schemes and require a training center to train the NNs.
Such CTDE schemes suffer from the scalability problem with the number of APs in the wireless network growing.
To address the above challenges, this article aims to develop a distributed-training-and-execution (DTE) power control algorithm based on MADRL for HetNets.

\subsection{Contributions}
We study the power control problem in a single-antenna multi-tier HetNet, where a macro AP is deployed in the center of the cell and several small-cell APs are deployed within the macro cell to enhance the performance of cell-edge UDs. 
A novel multi-agent penalty-based deep Q learning scheme is proposed to carry out the DTE power control scheme for HetNet. Our main contributions can be summarized as follows:
\begin{itemize}
	\item We propose a MADRL-based DTE power control scheme for HetNets, where each AP acts as an agent. 
	Specifically, at each AP, a deep Q network (DQN) is deployed to determine the transmit power independently and simultaneously with some exchanged information from neighboring APs.
	The proposed learning framework for HetNets can be trained and implemented in a distributed manner and thus has good scalability.
	\item We propose a novel penalty-based Q learning (PQL) for multi-agent reinforcement learning (MARL) systems to promote cooperation among agents.
	In PQL, each agent tends to choose an experienced action with high reward when revisiting a state, and thus the policy updating speed slows down.
	Therefore, it is easier for other agents to learn the policy of this agent and thus results in a more efficient collaboration process.
	Compared with the most basic MARL algorithm, i.e., independent Q learning (IQL) \cite{tan1993multi}, PQL only adds a few regularization terms in the loss function. 
	Hence, there is almost no extra cost to implement PQL than the basic IQL in a MARL system.
	We develop both Q table based and DQN-based PQL for general MARL problems.
	\item We implement our proposed PQL algorithm in the considered HetNet and compare it with other DTE MARL algorithms.
	Simulation results show that our proposed PQL can learn the desired power control policy from a dynamic environment where the locations of UDs change episodically, resulting in a better performance compared with IQL and hysteretic Q learning (HQL) \cite{matignon2007hysteretic}.
\end{itemize}

\subsection{Organization}
The remaining parts of this article are organized as follows.
Section \ref{system_model} introduces the considered HetNet system model and Section \ref{problem} formulates the sum-throughput maximization problem.
Our proposed PQL algorithm for multi-agent systems is introduced in Section \ref{PQL_intro}.
Our proposed DTE power control scheme based on PQL is then introduced in Section \ref{distributed_power_control}.
Section \ref{simulation_results} discusses the simulation results. Finally, Section \ref{conclusion} concludes the paper.


\section{System Model and Problem Statement}\label{system_model}
\begin{figure}
	\centering
	\includegraphics[width=0.65\textwidth]{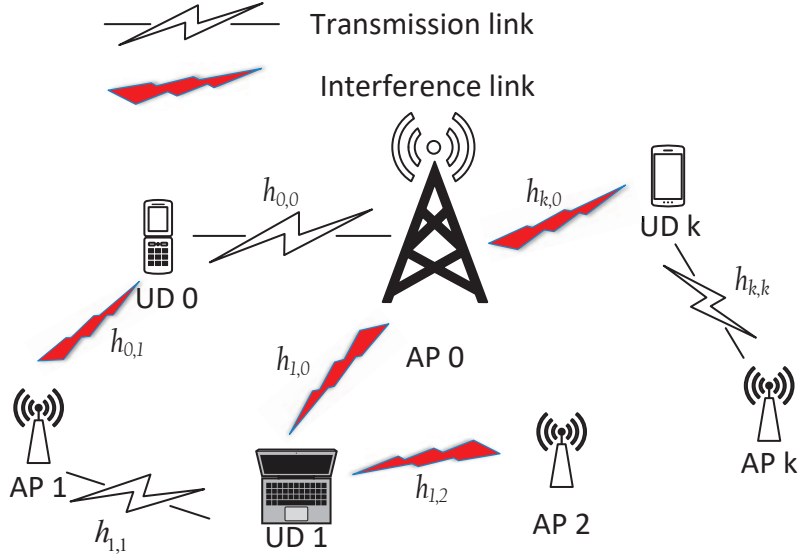}
	\caption{HetNet with two different tiers, which includes a macro-cell AP and multiple small-cell APs}
	\label{sys_diag}
\end{figure}
We consider a distributed power control problem in a single-antenna HetNet as illustrated in Fig. \ref{sys_diag}.
Within a macro cell, several small cells are deployed to offload transmission traffic or to enhance coverage.
There are $K$ APs and $K$ UDs denoted as $k\in\mathcal{K}=[0,\ldots,K-1]$, where AP $0$ is the macro AP, and AP $k$ serves UD $k$. 
All APs within the macro cell share the same spectrum to serve the $K$ UDs and may cause interference to others.
We consider a synchronized system and all channels are modeled as block fading as in \cite{ge2020deep}.
The channel gain, including transmission-link channel gain and interference channel gain, between AP $k$ and UD $m$ at time $t$ is modeled as,
\begin{equation}
	h_{m,k}^t = \sqrt{\beta_{m,k}}g_{m,k}^t, \forall m,k \in \mathcal{K},
\end{equation} 
where $g_{m,k}^t\in\mathbb{C}$ is the corresponding small-scale Rayleigh fading and $\beta_{m,k}$ is the large-scale fading, which is determined by the path-loss and log-norm shadowing and remains unchanged over a long time period.
We use Jakes fading model \cite{kim2011does} for the small-scale Rayleigh fading update, which can be expressed as follows,
\begin{equation}
	g_{m,k}^t = \rho g_{m,k}^{t-1}+ \sqrt{1-\rho^2} i_{m,k}^t, \forall m,k,\in \mathcal{K} \label{small_scale}
\end{equation}
where $i_{m,k}^t, \forall m, k \in\mathcal{K}$ are independent and identically distributed (i.i.d.) circularly symmetric complex Gaussian random variables with unit variance.
The correlation coefficient is $\rho = J_0(2\pi f_D T)$, where $J_0(.)$ denotes the $0$-th order Bessel function of first kind and $f_D$ and $T$ are the maximum Doppler frequency shift and the time slot duration, respectively.

At time slot $t$, AP $k$ transmits signal to user $k$ with transmit power $P_k^t$.
Assume the signal transmitted to user $k$ is $s_{k}^t$, then the received signal at user $k$ can be expressed as,
\begin{equation}
	y_{k}^t = \underbrace{h_{k, k}^{t}\sqrt{P_{k}^t}s_{k}^t}_{\text{Desired signal}}+\underbrace{\sum_{k'\neq k}h_{k, k'}^{t}\sqrt{P_{k'}^t}s_{k'}^t}_{\text{Interference}}+\,n_{k}, \label{transmission_sig}
\end{equation}
where $n_k$ denotes the additive white Gaussian noise (AWGN) with a variance of $\sigma_k^2$.
In the right side of \eqref{transmission_sig}, the first term is the desired signal, while the second term is the interference term, which includes both cross-tier and co-tier interference.
Therefore, the corresponding signal-to-interference-plus-noise ratio (SINR) at user $k$ can be expressed as,
\begin{equation}
	\gamma_{k}^t(\mathbf{P}^t) = \frac{|h_{k, k}^{t}|^2P_k^t}{\sum_{k'\neq k}|h_{k, k'}^{t}|^2P_{k'}^t  +\sigma_k^2},
\end{equation}
where $\mathbf{P}^t = \{P_{k}^t\mid\forall k\in\mathcal{K}\}$ denotes a collection of transmit powers from all APs.
The achievable rate of UD $k$ can thus be expressed as
$
r_{k}^t(\mathbf{P}^t) = \text{log}(1+\gamma_{k}^t(\mathbf{P}^t)).
$

\section{Problem Description}\label{problem}

We aim to maximize the system sum-throughput via optimizing transmit power of all APs in each time slot $t$.
This power control problem can be formulated as follows,
\begin{equation}\label{opt_prob}
	\begin{split}
		\max_{\mathbf{P}}\quad & \sum_{k\in\mathcal{K}} r_{k}^t(\mathbf{P}^t),\\
		\text{s.t.}\quad & P_{k}\leq P_{k}^{\mathrm{max}}, \forall k\in\mathcal{K},
	\end{split}
\end{equation}
where $P_{k}^{\mathrm{max}}$ is the maximum transmit power budget of AP $k$. 

In HetNet, different-tier APs have different transmit power budget, i.e., different $P_k^{\mathrm{max}}$ for each AP $k$, and observe the environments with different statistics, which makes it more difficult for agents to learn the desired power control policy.
In addition, the dynamic wireless environment with rapidly varying channels requires each AP to make decisions based on partial CSI only and to adjust transmit power before the data transmission stage starts so that the transmit power can maximize the system sum-throughput in every time slot.
However, the interference CSI $h_{m,k}^t,\forall m,k\in\mathcal{K},m\neq k$ is usually difficult to estimate in real systems, which is crucial when solving problem \eqref{opt_prob} by conventional algorithms.
Therefore, traditional optimization based power control algorithms \cite{shi2011iteratively, shen2018fractional} fail to be implemented in real-time systems due to the lack of global CSI and their high computational complexity.
Although DL-based methods can be trained to predict transmit power based on only partial CSI in a low-complexity manner, the lack of interference CSI leads to the lack of ground truth and gradients of the objective function, and thus limiting the effectiveness of both supervised learning \cite{sun2018learning} and unsupervised learning \cite{liang2019towards} methods, if considering the training phase of the DL-based methods.
Finally, the scalability problem should be taken into account due to the huge number of APs in real systems.
It is thus difficult for centralized DL-based methods to obtain the data and to learn the optimal policy in large-scale wireless systems.
Therefore, it is necessary to develop DTE MADRL-based power control algorithms, which will be present in detail in the following sections, to effectively solve the aforementioned challenges.

\section{Multi-agent Reinforcement Learning Systems}\label{PQL_intro}
A partially observable MARL system can be described as a partially observable Markov decision process (POMDP), which is defined by a tuple $<N, \mathbf{s}_t, \mathbf{a}_t^{(k)}, R_{t}^{(k)}, \mathbf{z}_{t}^{(k)}, P, O>$.\footnote{When involving POMDP, researchers usually use a concept "belief" to describe the agent's belief about the invisible environment state. In this article, we assume each agent's observation can well reflect its belief. We thus only use observation to stand for each agent's local state.}
Specifically, for a MARL system with $N$ agents, in each time step $t$, each agent $k$ observes the environment in state $\mathbf{s}_t$ and obtains a local observation of the environment $\mathbf{z}_t^{(k)} = O(\mathbf{s}_t, k)$, where $O(.)$ is the observation function mapping the environment state $\mathbf{s}_t$ to the specific observation $\mathbf{z}_t^{(k)}$ of agent $k$.
Then, based on the local observation $\mathbf{z}_t^{(k)}$, agent $k$ chooses its action $\mathbf{a}_t^{(k)}$, which is a part of the system joint action $\mathbf{a}_t$.
All agents implement the joint action $\mathbf{a}_t$ and interact with the environment.
Each agent $k$ receives a local reward $R_t^{(k)}$ from the environment and the environment state evolves to next state $\mathbf{s}_{t+1}$ with transition probability $P(\mathbf{s}_{t+1}\mid\mathbf{s}_t, \mathbf{a}_t)$.
Thereafter, each agent $k$ receives a new observation of the environment, $\mathbf{z}_{t+1}^{(k)}= O(\mathbf{s}_{t+1}, k)$.

In general, the MARL-powered wireless systems can be formulated as cooperative and endless games.\footnote{Although the interference among different cells makes APs perform like a competitive manner, in most cases, we have a network-wise key performance indicator (KPI) to capture the system performance. Therefore, all APs, in fact, cooperate together to maximize the system KPI.}
In this article, we focus on cooperative games, where all agents share the same reward or they have a common potential function \cite{chew2016potential}.
We promote cooperation by sacrificing the accuracy of Q value of some sub-optimal actions.
To illustrate, we start with the basic learning scheme in MARL, IQL\cite{tan1993multi}.

\subsection{Independent Q Learning}
IQL is a direct extension from single-agent reinforcement learning to MARL, where each agent treats other agents as a part of environment.
Therefore, each agent needs to learn both the environment transition and the policy of other agents.
Each agent $k$ aims to learn an optimal policy $\pi^{(k)}$ that maximizes the cumulative discounted reward,
\begin{equation}
	V_{t,\pi^{(k)}}^{(k)}(\mathbf{s}_t)=R_t^{(k)}+\sum_{j=1}^\infty \gamma^{j}R_{t+j}^{(k)},
\end{equation}
where $\gamma$ is the discount factor.

In the traditional IQL, each agent maintains a Q table to record the Q values of all observation-action pairs of the agent.
The Q values are learned from the experiences of the agents interacting with the environment.
When receiving an experience tuple $\mathbf{e}_t^{(k)}= \,<\mathbf{z}_t^{(k)}, \mathbf{a}_t^{(k)}, R_t^{(k)}, \mathbf{z}_{t+1}^{(k)}>$, agent $k$ updates its Q table as follows,
\begin{equation}
	Q_k(\mathbf{z}_t^{(k)}, \mathbf{a}^{(k)}_t) \leftarrow Q_k(\mathbf{z}_t^{(k)}, \mathbf{a}_t^{(k)})+\alpha \delta_t^{(k)},
\end{equation}
where $\delta_t^{(k)} = R_t^{(k)} + \gamma\max_{\mathbf{a}^{*(k)}} Q_k(\mathbf{z}_{t+1}^{(k)}, \mathbf{a}^{*(k)})- Q_k(\mathbf{z}_{t}^{(k)}, \mathbf{a}_t^{(k)})$ is the temporal difference (TD) error.
$\alpha$ is the learning rate.

In IQL, each agent learns its Q table based on its own experience in a distributed manner with other agents' policies treated as a part of the environment.
However, since all agents keep updating their policies simultaneously, the environment observed by one agent is no longer static, which violates the assumption of Q learning that the environment state transition probability matrix is fixed.
The varying policies of other agents make it difficult for the agent to learn the environmental dynamics and leads to instability and performance degradation of IQL.

%
\subsection{Proposed Penalty-based Q Learning for MARL Systems}\label{q_table_pql}
One major challenge of MARL in wireless communication systems is that the reward structure is very complex.
Many different resource allocation solutions have similar performance, which results in the fact that at a specific observation, the Q values of different actions are close to each other.
On the other hand, for each agent, the covariance of Q value for a specific observation-action pair increases because of the uncertainty of other agents' policies.
Therefore, during the training phase, the update of Q table can easily lead to the change of the action with the maximum Q value given an observation.
This can result in the rapid change of the policy and thus make it harder for agents to learn to cooperate.
We promote the cooperation by slowing down the policy updating speed and focusing on the experiences with high reward.
That is to say, during the training phase, agents will tend to choose an experienced action with high reward when revisiting an observation and thus learn the Q values of these high-reward observation-action pairs first.

In RL systems, the policy of an agent is given by $\pi^{(k)}=\argmax_{\mathbf{a}^{(k)}} Q_k(\mathbf{z}_t^{(k)}, \mathbf{a}^{(k)})$, which means the selected action is determined by the maximum Q value given an observation $\mathbf{z}_t^{(k)}$.
Therefore, in our proposed PQL, we sacrifice the accuracy of non-maximum observation-action Q values to let the agent stay on observation-action pairs with high reward first.
Specifically, we modify the learning steps of the IQL.
When receiving an experience tuple $ \mathbf{e}_t^{(k)}$, instead of updating the Q value of one observation-action pair, PQL updates the Q values of all actions given the observation $\mathbf{z}_t^{(k)}$. The PQL updating rules are expressed as in \eqref{Q_update},
\begin{figure*}
	\begin{equation}\label{Q_update}
		Q_k(\mathbf{z}_t^{(k)}, \mathbf{a}^{(k)})\!\leftarrow\! \begin{cases}
			Q_k(\mathbf{z}_t^{(k)}, \mathbf{a}^{(k)})+\alpha \delta_t^{(k)}, &\!\!\!\!\text{if } \mathbf{a}^{(k)} = \mathbf{a}_t^{(k)},\\
			Q_k(\mathbf{z}_t^{(k)}, \mathbf{a}^{(k)}) - \beta, &\!\!\!\!\!\!\!\text{if } \mathbf{a}^{(k)}\neq\mathbf{a}_t^{(k)} \text{and } C_1^k(\mathbf{e}_t^{(k)})>t_1 \text{and } C_2^k(\mathbf{z}_{t}^{(k)},\mathbf{a}^{(k)})<t_2,
		\end{cases}
	\end{equation}
\end{figure*}
where $\delta_t^{(k)}$ is the corresponding TD error. 
$\beta$ is the penalty factor.
$C_1^k(\mathbf{e}_t^{(k)})$ and $C_2^k(\mathbf{z}_{t}^{(k)},\mathbf{a}^{(k)})$ are conditional variables, whose expressions are as follows,
\begin{equation}
	C_1^k(\mathbf{e}_t^{(k)})=R_t^{(k)}+\gamma\max_{\mathbf{a}^{*(k)}} Q_k(\mathbf{z}_{t+1}^{(k)}, \mathbf{a}^{*(k)}) - \max_{\mathbf{a}^{*(k)}}Q_k(\mathbf{z}_{t}^{(k)}, \mathbf{a}^{*(k)}),
\end{equation}
\begin{equation}
	C_2^k(\mathbf{z}_{t}^{(k)},\mathbf{a}^{(k)}) = \max_{\mathbf{a}^{*(k)}} Q_k(\mathbf{z}_{t}^{(k)}, \mathbf{a}^{*(k)})-Q_k(\mathbf{z}_{t}^{(k)}, \mathbf{a}^{(k)}).
\end{equation}
Specifically, $C_1^k(\mathbf{e}_t^{(k)})$ is used to determine whether the received experience tuple is "significantly good", and $C_2^k(\mathbf{z}_{t}^{(k)},\mathbf{a}^{(k)})$ is used to select those actions with estimated Q values close to the maximum Q value at the current observation $\mathbf{z}_{t}^{(k)}$.
$t_1>0$ and $t_2>0$ are two thresholds.

With the PQL updating rules in \eqref{Q_update}, when an agent $k$ receives an experience tuple $\mathbf{e}_t^{(k)}$ with significantly high reward, which is triggered by the condition $C_1^k(\mathbf{e}_t^{(k)})>t_1$, it degrades the Q values of other unselected actions with Q value close to the maximum one, which are filtered out by the condition $\mathbf{a}^{(k)}\neq\mathbf{a}_t^{(k)} \text{and } C_2^k(\mathbf{z}_{t}^{(k)},\mathbf{a}^{(k)})<t_2$. 
Therefore, when agent $k$ revisits this observation, the previously-visited high-reward observation-action is more likely to remain the highest Q value among all actions in this observation due to the Q value degradation to other actions in previous visits although the agent has updated its Q table for many steps.
As a result, the agent tends to choose the experienced action with high reward and thus the policy updating speed slows down.
After the agent has well learned the high-reward observation-action pairs, the condition $C_1^k(\mathbf{e}_t^{(k)})>t_1$ is no longer satisfied and then the PQL turns back into IQL, which results in the exploration of other observation-action pairs.\footnote{
We employed this Q table based PQL to play the two-agent predator-prey game as described in \cite{matignon2007hysteretic}.
The simulation results show the superiority of PQL to the hysteretic q learning from \cite{matignon2007hysteretic}.
For the consistence of this article, we omit this experiment in this article, but we will publish the simulation code of the predator-prey game together with that of the experiments in HetNet after the acceptance of this article.}

\subsection{Penalty-based Deep Q Learning}
In wireless multi-agent systems, especially in the resource allocation scenarios, the state is usually determined by the system global CSI, which is continuous and not suitable for traditional Q table based RL algorithms.
As a consequence, to solve the power control problem in HetNet, we need to extend our PQL to deep Q learning scenarios.

In the context of deep Q learning, the Q table is replaced by the DQN, which estimates the Q values of all actions with a given observation.
Therefore, we can realize the PQL in deep Q learning by penalizing the outputs of the DQN corresponding to the unselected actions.
Specifically, we add a few regularization terms in the Q loss function.
The resulted loss function for agent $k$ of our multi-agent deep PQL algorithm is given by \eqref{loss},
\begin{figure*}
	\begin{equation}
		L(\mathbf{e}_t^{(k)};\boldsymbol{\theta}_t^{(k)}) = (\delta_t^{(k)})^2 + \beta\sum_{\mathbf{a}^{(k)}\neq\mathbf{a}_t^{(k)}}C_{1}^{k}(\mathbf{e}_t^{(k)})C_{2}^k(\mathbf{z}_{t}^{(k)},\mathbf{a}^{(k)})Q_k(\mathbf{z}_{t}^{(k)},\mathbf{a}^{(k)};\boldsymbol{\theta}_t^{(k)}),
		\label{loss}
	\end{equation}
\end{figure*}
where $\beta$ is the penalty factor, $Q_k(.;\boldsymbol{\theta}_t^{(k)})$ denotes the DQN of agent $k$ at time slot $t$ with respect to the network parameters $\boldsymbol{\theta}_t^{(k)}$.
$\delta_t^{(k)}=R_t^{(k)} + \gamma\max_{\mathbf{a}^{*(k)}} Q_k^{\mathrm{tar}}(\mathbf{z}_{t+1}^{(k)}, \mathbf{a}^{*(k)};\boldsymbol{\theta}^{(k)-})- Q_k(\mathbf{z}_{t}^{(k)}, \mathbf{a}_t^{(k)};\boldsymbol{\theta}_t^{(k)})$ denotes the TD error of agent $k$ at time slot $t$, where $Q_k^{\mathrm{tar}}(.;\boldsymbol{\theta}^{(k)-})$ is the target DQN and $\boldsymbol{\theta}^{(k)-}$ is the corresponding target network parameters.
As presented in Section \ref{q_table_pql}, $C_{1}^{k}(\mathbf{e}_t^{(k)})$ and $C_{2}^k(\mathbf{z}_{t}^{(k)},\mathbf{a}_t^{(k)})$ are the conditional variables given in \eqref{C1_con} and \eqref{C2_con}, respectively,
\begin{figure*}
	\begin{equation}\label{C1_con}
		C_{1}^{k}(\mathbf{e}_t^{(k)})=\begin{cases}
			1, &\text{if } R_t^{(k)} + \gamma\max_{\mathbf{a}^{*(k)}} Q_k^{\mathrm{tar}}(\mathbf{z}_{t+1}^{(k)}, \mathbf{a}^{*(k)};\boldsymbol{\theta}^{(k)-})\\ &\quad\quad -\max_{\mathbf{a}^{*(k)}}Q_k(\mathbf{z}_{t}^{(k)}, \mathbf{a}^{*(k)};\boldsymbol{\theta}_t^{(k)}) > t_1\\
			0, &\text{otherwise,}
		\end{cases}
	\end{equation}
	\begin{equation}\label{C2_con}
		C_{2}^{k}(\mathbf{z}_{t}^{(k)},\mathbf{a}^{(k)})=\begin{cases}
			1, &\text{if } \max_{\mathbf{a}^{*(k)}} Q_k(\mathbf{z}_{t}^{(k)}, \mathbf{a}^{*(k)};\boldsymbol{\theta}_t^{(k)})-Q_k(\mathbf{z}_{t}^{(k)}, \mathbf{a}^{(k)};\boldsymbol{\theta}_t^{(k)})<t_2\\
			0, &\text{otherwise,}
		\end{cases}
	\end{equation}
\end{figure*}
where $t_1>0$ and $t_2>0$ denote the corresponding thresholds.

\section{Proposed DTE Power Control Scheme}\label{distributed_power_control}
In this section, we implement our proposed PQL to allocate transmit power for the HetNet, in which each AP acts as an agent, which interacts with the environment and gains the experience for training.
Each agent needs to gradually adjust its own power control strategy based on its local experience.
We design a local reward for each agent, which considers the achievable rates of nearby cells, so that the whole multi-agent system works in a cooperative manner in the interest of global network performance.

Our proposed MARL scheme has two phases, including training phase and implementation phase.
In this article, we focus on DTE power control schemes.
In such DTE schemes, each agent maintains a local DQN and updates the DQN based on its local experience to maximize its local reward. 
In addition, each agent decides its transmit power based on the collected local observation independently and simultaneously on a time scale on par with the small-scale channel fading.

\subsection{State and Observation Space}

APs are allowed to exchange some information before each transmission stage. 
At the beginning of each time slot, each AP collects local measurements from its neighbors via wired or wireless backhaul links \cite{haija2017small}.
The local measurement of AP $k$, $\mathcal{O}_k^t$, includes the achievable rate, $r_k^{t-1}(\mathbf{P}^t)$, and the received interference plus noise power $I_k^{t-1}=\sum_{k'\neq k} h_{k, k'}^{t-1}P_{k'}^{t-1}+\sigma_k^2$ measured in the last time slot $t-1$, i.e., $\mathcal{O}_k^t=\{r_k^{t-1},I_k^{t-1}\}$.
Therefore, together with the estimated transmission link channel gain, $h_{k,k}^t$, the local observation of AP $k$ can be expressed as $\mathbf{z}_t^{(k)} = \{h_{k,k}^t,\mathcal{O}_{k'}^t\mid k'\in\mathcal{N}_k\cup\{k\}\}$, where $\mathcal{N}_k$ denotes the index set of the nearby cells. 
The neighboring set of AP $k$, $\mathcal{N}_k$, is the index set of $M$ UDs served by other APs with the smallest distance to AP $k$.
As a result, each agent has a local observation of $2M+3$ dimensions.

Note that the local observation $\mathbf{z}_t^{(k)}$ only includes the transmission link channel gain and the local measurements of itself and its neighbors.
Therefore, each AP only needs to broadcast two scalars to its neighbors in each time slot.
We thus assume that the local observation can be obtained without delay at the beginning of each time slot.
In addition, the local reward we design in Section \ref{reward_design} does not require any additional information beside the local observation we discuss above.

\subsection{Action Space}
In this article, we limit the transmit power to discrete levels for ease of learning and practical circuit restriction.
Specifically, each AP $k$ with maximum transmit power budget $P_{k}^{\mathrm{max}}$ has a discrete power level set of $[0, \frac{P_{k}^{\mathrm{max}}}{A_C-1}, \frac{2P_{k}^{\mathrm{max}}}{A_C-1}, ..., P_{k}^{\mathrm{max}}]$, where $A_C$ is the cardinality of the action set. 
Therefore, each agent has an action space of $A_C$, and each action refers to a specific transmit power level in the action space.

\subsection{Reward Design}\label{reward_design}
One advantage of RL is the flexibility of reward design. 
When the reward is correlated to the desired objective, the system performance can be increased.
Our goal is to improve the system sum-throughput and train the MARL system locally.
Therefore, we need to design a reward obtained from local information and related to the system sum-throughput for each agent.
To do so, we use the achievable rates obtained from neighbors and define the reward for time slot $t$ at agent $k$ as follows,
\begin{equation}
	R_{k}^t=\frac{1}{M+1}\sum_{k'\in\mathcal{N}_k\cup\{k\}}r_{k'}^{t}. \label{reward}
\end{equation}
Since each AP's transmit power mainly affects the achievable rates of its neighboring UDs, the local reward defined in \eqref{reward} can successfully characterize the system sum-throughput.

\subsection{Distributed Implementation Phase}
The distributed power control algorithm is summarized in Algorithm \ref{Power control implementation}, where we set $is\_train=False$.
Specifically, in the implementation phase, at time slot $t$, each agent $k$ first estimates and collects its local measurement, i.e.,  $h_{k,k}^t$, $r_k^{t-1}$ and $I_k^{t-1}$ as in Step 3.1.
Then it exchanges the local measurement with its neighbors and forms the local observation $\mathbf{z}_t^{(k)}$ as in Step 3.2.
Based on local observation $\mathbf{z}_t^{(k)}$, agent $k$ selects the action $a_k^t$ corresponding to the maximum Q value as in Step 3.4.
All APs then start transmission with the transmit power corresponding to the selected actions as in Step 3.5.

Note that the information exchange can be done before next time slot starts once the SINR and interference power are estimated with no delay.
In addition, each AP only needs to implement a local low-complexity forward propagation to adjust their transmit power.
As a result, our proposed MADRL-based power control algorithm can make decision on time and thus is more practical compared with optimization-based algorithms.

\subsection{Decentralized Training Phase}
\begin{algorithm}
	\caption{Proposed power control scheme}
	\begin{itemize}
		\item[1.] Episode $m$ starts. Users are randomly located.
		\item[2.] Time slot $t$ starts. Small scale channel gain is updated according to \eqref{small_scale}.
		\item[3.] For each AP $k$,
		\begin{itemize}
			\item[3.1] Before updating the power, user $k$ estimates its direct link channel gain $h_{k,k}^t$ and the interference plus noise power strength $I_i^{t-1}$, as well as the achievable rate of the last time slot $r_k^{t-1}$, and then feeds them back to AP $k$.			
			\item[3.2] Each AP $k$ obtains the local measurements $\mathcal{O}_{j}^t$ from its neighboring APs $j\in\mathcal{N}_k$ through backhaul links and then forms the local observation $\mathbf{z}_t^{(k)}$.
			\item[3.3] If $is\_train==True$:
			\begin{itemize}
				\item[3.3.1] Each AP $k$ forms the experience tuple $(\mathbf{z}_{t-1}^{(k)},a_k^{t-1}, R_k^{t-1})$ and stores it in its local replay buffer.
				\item[3.3.2] Each AP $k$ uniformly samples a mini-batch $\mathcal{D}_k$ from the replay buffer.
				\item[3.3.3] Each AP $k$ updates its local DQN by minimizing the loss function \eqref{loss} on the mini-batch $\mathcal{D}_k$ using variant of stochastic gradient
				descent.
			\end{itemize}
			\item[3.4] If $is\_train==True$:
			\begin{itemize}
				\item[] AP $k$ predicts its transmit power, $\hat{P}_k^t$, via $\epsilon$-greedy, i.e., to select the action corresponding to the highest Q value with a probability of $1-\epsilon$ or to randomly choose its transmit power with a probability $\epsilon$.
			\end{itemize}
			\item[] else:
			\begin{itemize}
				\item[] AP $k$ predicts its transmit power, $\hat{P}_k^t$, based on the local observation $\mathbf{z}_t^{(k)}$ by selecting the action corresponding to the highest Q value.
			\end{itemize}										
			\item[3.5] AP $k$ adjusts the transmit power as $P_k^t = \hat{P}_k^t$.
		\end{itemize}				
		\item[4.] Time slot $t$ ends. Time slot index $t=t+1$. Go back to Step 2 if episode $m$ does not end.
		\item[5.] Episode $m$ ends.	Episode index $m=m+1$. Go back to Step 1.
	\end{itemize}
	
	\label{Power control implementation}
\end{algorithm}

We focus on the training in a dynamic episodic setting where users' locations are fixed during each episode.
When an episode begins, the locations of all users are regenerated uniformly in the corresponding cells and the initial transmit powers of all cells are set to $P_{k}^{\mathrm{max}}, \forall k\in\mathcal{K}$.
The large-scale fading is then generated based on the users' locations.
The small-scale fading is updated when a time slot begins. 
The change of small-scale channel fading triggers a transition of the environment state and causes each agent to adjust its transmit power.

We leverage deep Q learning with experience replay \cite{mnih2015human} to train the HetNet agents for effective learning of local power control policies.
The training phase is summarized in Algorithm \ref{Power control implementation}, where $is\_train=True$.
Compared with the implementation phase, the training phase has two different steps, Step 3.3 and Step 3.4.
Step 3.3.1 is the experience collecting step, where all APs store their local experience tuples of the last time slot in their local replay buffers.
Steps 3.3.2 and 3.3.3 are the network training steps, where each AP $k$ uniformly samples a mini-batch data set $\mathcal{D}_k$ and then updates the trainable parameters of its DQN with the stochastic gradient descent method to minimize the PQL loss function as defined in \eqref{loss}.
Step 3.4 is the decision making step.
In the training mode, each agent chooses its action according to the widely-used epsilon-greedy strategy, i.e., to randomly choose an action with probability of $\epsilon$, or to choose the action corresponding to the highest Q value given the current observation with probability of $1-\epsilon$.

Note that we can easily change the simplest MADRL training algorithm, IQL, into our PQL easily.
The only extra cost comes from the calculation of the added regularization terms.
However, in each backward propagation, these regularization terms only involve simple linear and comparison operations and there is no extra computational complexity in back propagating from the output layer to each trainable parameters of the NN.
Therefore, PQL only introduces very limited computational complexity of calculating the gradients of the regularization terms during the training phase.
As a result, our proposed PQL algorithm has practical distributed training phase due to its IQL-like training manner.

\subsection{Network Structure}
\begin{figure}
	\centering
	\includegraphics[width=0.65\textwidth]{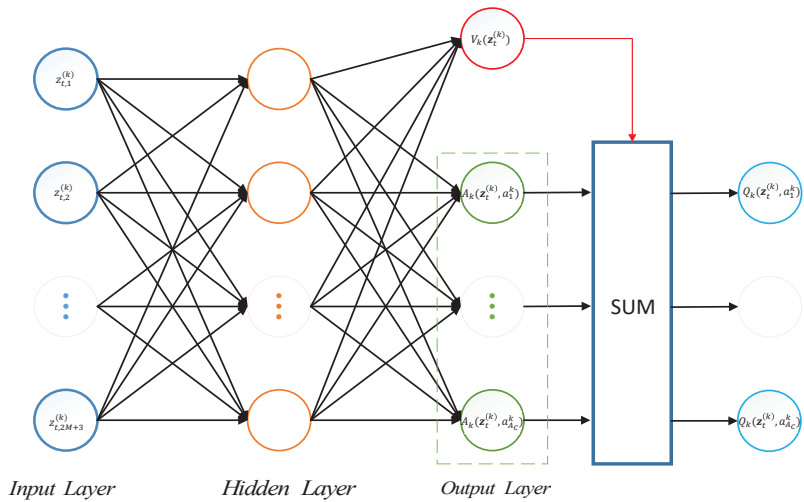}
	\caption{Network structure of dueling DQN.}
	\label{dueling_dqn}
\end{figure}
To further improve the performance without imposing any change to the underlying reinforcement learning algorithm, we employ the deep dueling DQN \cite{wang2016dueling} structure for our agents.
Specifically, the network structure of dueling DQN is illustrated in Fig. \ref{dueling_dqn}.
In dueling DQN, the Q value of an observation-action\footnote{As mentioned before, we use observation to indicate the local state for each agent. Therefore, we use the term "observation-action" instead of "state-action" in this article.} pair is decomposed as the sum of the value function and the advantage function, which is expressed as follows,
\begin{equation}\label{dueling}
	Q_k(\mathbf{z}_t^{(k)},\mathbf{a}^{(k)}) = V_k(\mathbf{z}_t^{(k)})+A_k(\mathbf{z}_t^{(k)},\mathbf{a}^{(k)}),
\end{equation}
where $V_k(\mathbf{z}_t^{(k)})$ is the value function, which reflects the average of Q values given the observation $\mathbf{z}_t^{(k)}$, and $A_k(\mathbf{z}_t^{(k)})$ is the advantage function, which measures the relative importance of a certain action
compared with other actions.
There is a zero-sum constraint on the advantage function $A_k(.)$, which is given by,
\begin{equation}
	\sum_{\mathbf{a}^{(k)}}A_k(\mathbf{z}_t^{(k)},\mathbf{a}^{(k)})=0,\forall \mathbf{z}_t^{(k)}.
\end{equation}
The intuition behind Dueling DQN is that it is not always necessary to estimate the value of taking each available action. 
For some states, the choice of action makes limited influence on what happens.
One the other hand, with the decomposition \eqref{dueling}, during the backward propagation, the NN tends to update value function $V_k(\mathbf{z}_t^{(k)})$ instead of the Q value of one observation-action pair, which results in the learning of all Q values of the same observation.
For the input and hidden layers in Fig. \ref{dueling_dqn}, we use the basic fully-connected structure for all agents.
Specific hyper-parameters regarding number of layers and neurons and types of activation function will be illustrated in the next section.


\section{Simulation Results}\label{simulation_results}

\begin{figure}
	\centering
	\includegraphics[width=0.65\textwidth]{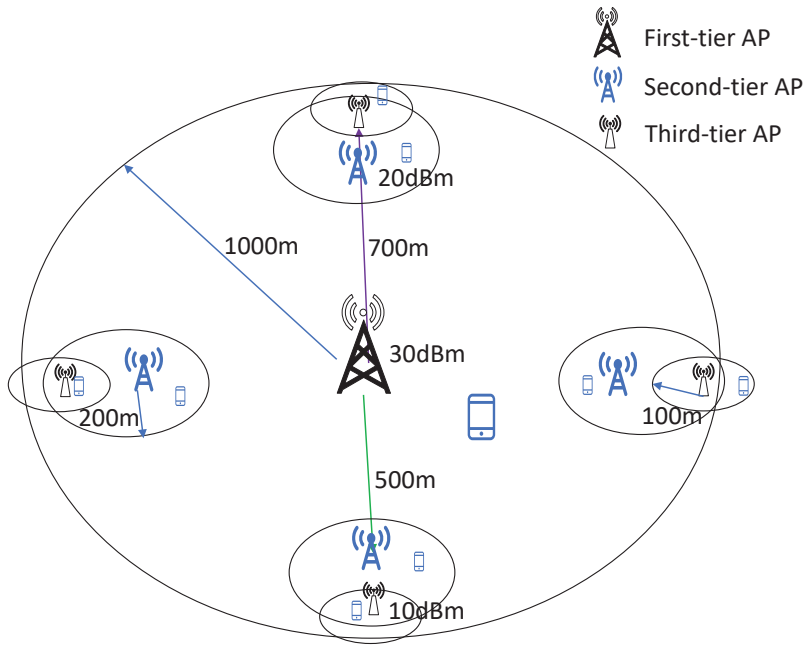}
	\caption{HetNet with three tiers.}
	\label{AP_depoy}
\end{figure}

In this section, we present simulation results to verify the effectiveness of our proposed distributed power control scheme based on MADRL in HetNets.
Unless otherwise stated, our HetNet system parameters are as the following.
We consider a three-tier HetNet as illustrated in Fig. \ref{AP_depoy}.
In this HetNet, there are nine APs in total, which are located at $(0,0)$m, $(500, 0)$m, $(0,500)$m, $(-500,0)$m, $(0,-500)$m, $(700,0)$m, $(0,700)$m, $(-700,0)$m and $(0, -700)$m, respectively, indexed by $[0,\ldots,8]$ \cite{zhang2020deep}.
AP $0$ is the macro-cell or first-tier AP with a coverage defined by a minimum distance, $d_{min}=10$m and a maximum distance, $d_{max}=1,000$m, from the AP to UDs.
APs $\{1,2,3,4\}$ are small-cell APs in the second tier whose coverage is defined by $d_{min}=10$m and $d_{max}=200$m.
APs $\{5,6,7,8\}$ are small-cell APs in the third tier whose coverage is defined by $d_{min}=10$m and $d_{max}=100$m.
UDs are uniformly located in their corresponding cells.
The maximum transmit power budget is set to $30$ dBm, $20$ dBm and $10$ dBm for the first-tier, second-tier and third-tier APs, respectively, over a fully reusable frequency band of 10 MHz.
Following the LTE standard \cite{TR25942}, the path loss is given by $120.9+37.6\log_{10}(d/1,000)$ (dB), where $d$ (m) is the distance between the transmitter and the receiver.
The standard deviation of the log-norm shadow fading is 8 dB.
The AWGN power is $-114$ dBm.
The time slot duration $T=20$ ms and we assume the same maximum Doppler frequency shift $f_D = 10$ Hz for all UDs \cite{nasir2019multi}.
The number of considered neighbors, $M$, is set as 4.
The number of discrete transmit power level is set as $A_C=11$.

We employ the fully connected dueling DQN structure for all agents.
The dueling DQN has two fully connected hidden layers with 128, 64 neurons, respectively.
The activation functions of the input layer and first two hidden layers are the Leaky ReLU function with the negative slope being 0.1, while the output layer does not have activation functions.
We set the memory size of the replay buffer as $D_{max} = 3,600$ and the number of time slots in each episode as $N_{slots} = 20$ so that the agent can learn the environmental dynamics caused by the change of large scale fading.
The mini-batch size is set to 128.
We use a decaying $\epsilon$-greedy exploration parameter, which can be expressed as a function of training steps $\epsilon = \max(1-0.7*1.0005^t,0.01)$.
Adam optimizer with learning rate of 0.0001 is used to train the dueling DQN.
Since we only aim to maximize the system sum-throughput in each time slot, the future reward is unimportant.
We thus set the discount factor $\gamma=0$.
The penalty factor of PQL, $\beta$, is set as 0.05, while the two thresholds are set as $t_1=0.1R_t^{(k)}$ and $t_2=1$, respectively.

We compare our proposed scheme with some other DTE learning based scheme, including WMMSE, IQL, HQL, and full power.
\begin{itemize}
	\item[1)]WMMSE: The weighted minimum mean-squared error (WMMSE) algorithm  \cite{shi2011iteratively} is centralized with full-CSI and should has the best performance.
	\item[2)]PQL: PQL is our proposed DTE power control scheme.
	\item[3)]IQL: Its settings, including its DQN hyper-parameters, are the same to our proposed PQL scheme, except for the loss function, which still follows the original IQL loss.
	\item[4)]HQL: Hysteretic Q-learning \cite{matignon2007hysteretic} uses two different learning rates to train the DQN. 
	When receiving an experience, the agent first compares the estimated Q value by its DQN with the target Q value, which is calculated by the sum of reward and the discounted Q value of the next step.
	If the target Q value is higher, it uses a normal learning rate to learn this experience, otherwise, it uses a degraded learning rate to learn this experience.
	We set the degraded learning rate as 0.4 as in \cite{omidshafiei2017deep}.
	\item[5)]Full power: Each AP transmits signal using maximum power.
\end{itemize}

To begin with, we compare our proposed method with other benchmarks in the considered scenario.
Specifically, to better evaluate the performance of the MARL system, we use two identical environments in out experiments, i.e., training environment and testing environment.
We train the agents in the training environment, which updates the trainable parameters of all agents, while test the performance in the testing environment with all trainable parameters fixed.
In the testing phase, the DQNs are used to optimize the transmit power at each AP for $4,000$ time slots, i.e., $200$ episodes.
Therefore, all the curves in the following figures are the average performance over $200$ episodes.

\begin{figure}
	\centering
	\includegraphics[width=0.65\textwidth]{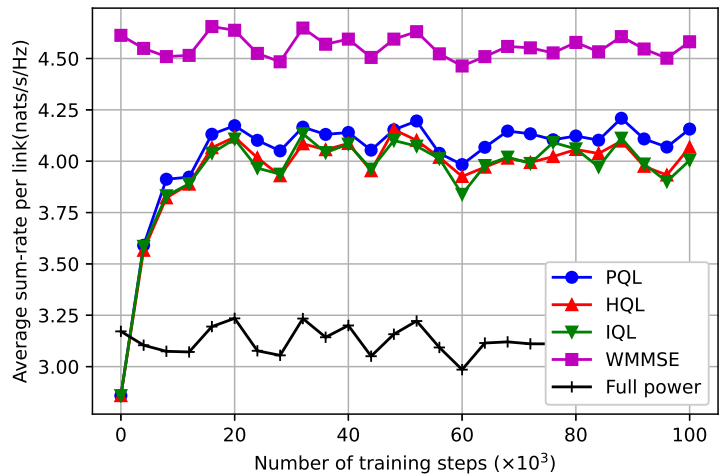}
	\caption{Average sum-throughput performance of different algorithms in the training phase.}
	\label{abs_alg}
\end{figure}
\begin{figure}
	\centering
	\includegraphics[width=0.65\textwidth]{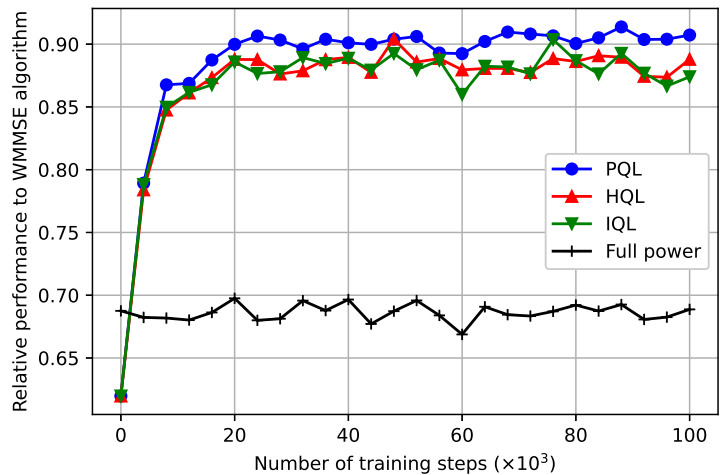}
	\caption{Relative performance of different algorithms in the training phase.}
	\label{rel_alg}
\end{figure}

Fig. \ref{abs_alg} shows the average sum-throughput performance of the
proposed algorithm and other benchmark algorithms in the investigated HetNet.
From the figure, all the learning based algorithms have worse performance than the full power scheme at the beginning since their DQNs are randomly initialized and thus randomly choose transmit power.
After training for $1,000$ episodes, i.e. $20,000$ time slots, the performance of MADRL-based algorithms comes to a stage.
In addition, our proposed PQL outperforms other MADRL-based methods all the time and the performance gain over IQL is around $3\%$.
This is because in PQL, each agent tends to choose an experienced action with high reward when revisiting a state, and thus the policy updating speed slows down. In this way, each agent's policy can be learned by other agents more easily, resulting in a more efficient cooperation-learning process.
Note that the UD locations are regenerated randomly in every episode, which causes the performance fluctuation to all algorithms.
Therefore, to better illustrate the performance of different learning methods and to show the effectiveness of these learning algorithms, we illustrate the relative performance of the considered methods to the WMMSE algorithm in Fig. \ref{rel_alg}.
We can observe that PQL can reach over $90\%$ of the sum-throughput achieved by the WMMSE algorithm.
In addition, Fig. \ref{rel_alg} also indicates that our proposed PQL has more stable performance compared with IQL algorithm during the training phase.

\begin{figure}
	\centering
	\includegraphics[width=0.65\textwidth]{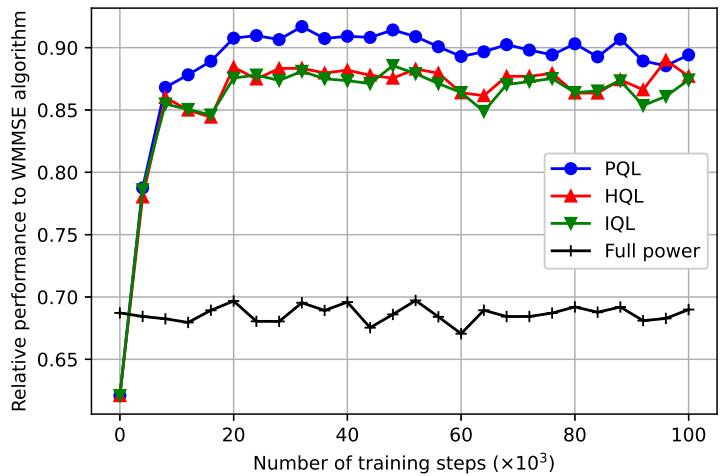}
	\caption{Relative performance of different algorithms in the training phase when $\rho=0$.}
	\label{zero_rho}
\end{figure}

When using MADRL methods, the local observation of each agent includes SINR and interference information regarding the CSI of previous time slot.
Therefore, when $\rho=0$, agents can obtain less information from its local observation about CSI of the current time slot.
We thus set the correlation coefficient $\rho=0$ and keep other system settings unchanged to evaluate the performance of these algorithms as shown in Fig. \ref{zero_rho}.
Our proposed PQL still outperforms IQL and HQL with performance gain of around $4\%$, although it also encounters performance degradation after 60,000 training steps.
This result validates the ability of our PQL to learn to cooperate given less state information.

\begin{figure}
	\centering
	\includegraphics[width=0.65\textwidth]{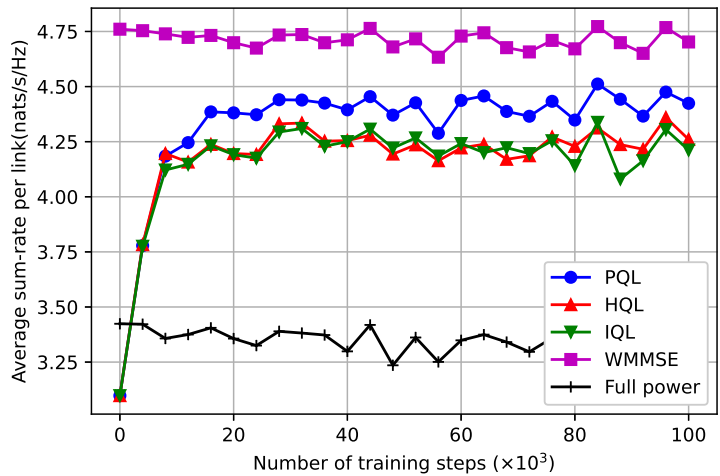}
	\caption{Average sum-throughput performance of different algorithms in the training phase in the HetNet with 4 additional third-tier cells.}
	\label{abs_algs_8f}
\end{figure}

\begin{figure}
	\centering
	\includegraphics[width=0.65\textwidth]{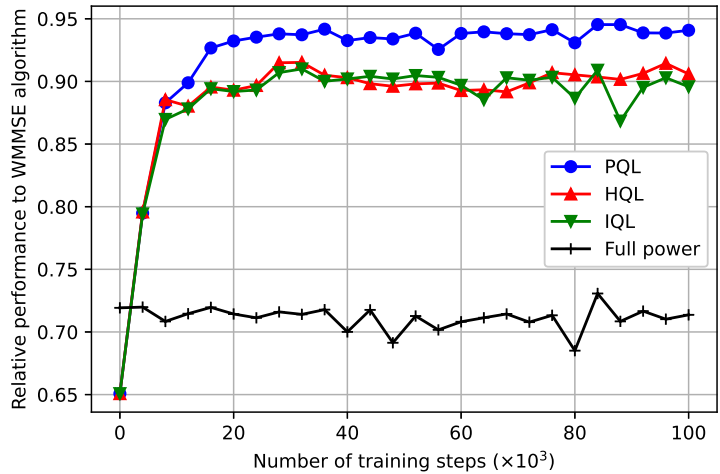}
	\caption{Relative performance of different algorithms in the training phase in the HetNet with 4 additional third-tier cells.}
	\label{8fcell}
\end{figure}

In the following experiment, we add 4 more third-tier cells at the edge of the macro cell to investigate PQL with more agents.
Specifically, the newly added third-tier APs are located at $(250\sqrt{2}, 250\sqrt{2})$m, $(250\sqrt{2}, -250\sqrt{2})$m, $(-250\sqrt{2}, -250\sqrt{2})$m and $(-250\sqrt{2}, 250\sqrt{2})$m, respectively, with other system parameters the same as other third-tier APs.
Fig. \ref{abs_algs_8f} and Fig. \ref{8fcell} draw the system sum-throughput performance of considered algorithms and the relative performance to the WMMSE algorithm of the considered algorithms during the training phase.
From Fig. \ref{abs_alg} and Fig. \ref{abs_algs_8f}, with more third-tier APs deployed at the macro-cell edge, the system average sum-throughput performance of all algorithms is increased.
This result validates the benefit of deploying small cells within macro cells.
In addition, from Fig. \ref{rel_alg} and Fig. \ref{8fcell}, the performance gain gap between PQL and IQL/HQL is around $5\%$ and increases with the number of agents, although the relative performance of all MADRL methods increases.
This is because with the number of agents increasing, better cooperation among agents can provide more performance gain.
Our proposed PQL can efficiently promote the cooperation among agents and thus obtains more performance gain than PQL and IQL.

\begin{figure}
	\centering
	\includegraphics[width=0.65\textwidth]{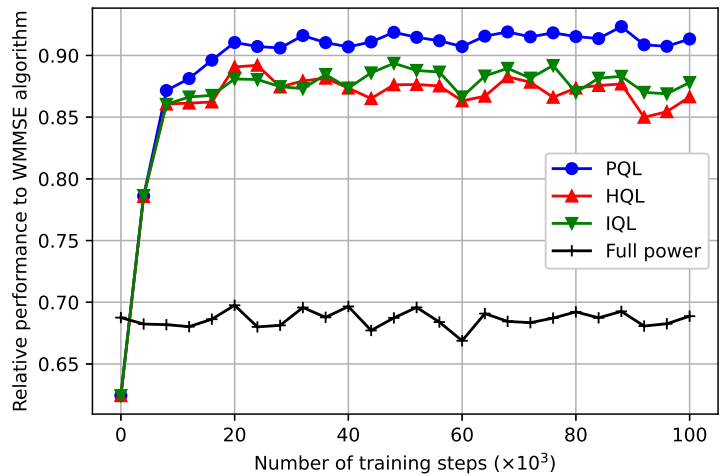}
	\caption{Relative performance of different algorithms in the training phase with different-structure DQNs deployed at different-tier APs.}
	\label{diffModel}
\end{figure}

Considering the difference of computational capability, battery capacity and environmental statistics of different-tier APs in HetNets, different-tier APs can be deployed with different-structure DQNs.
Therefore, in the last experiment, we use different-structure dueling DQNs in different-tier APs to investigate the ability of our PQL to train heterogeneous multi-agent systems.
For the first-tier AP, its dueling DQN has three hidden layers with 128, 128, 64 neurons, respectively.
For second-tier APs, their dueling DQNs have two hidden layers with 128, 64 neurons, respectively.
For third-tier APs, their dueling DQNs have two hidden layers with 64, 64 neurons, respectively.
The resulted relative performance during the training phase is shown in Fig. \ref{diffModel}.
Our proposed PQL algorithm reaches over $90\%$ sum-rate of the WMMSE algorithm and outperforms other MADRL algorithms as expected while HQL performs worse than IQL.
In this scenario, our proposed PQL has a performance gain of around $4\%$ over IQL.
In addition, PQL has more stable performance than HQL and IQL after $20,000$ time slots.
This result validates our PQL on its ability of training a MADRL system with heterogeneous agents.

\section{Conclusions}\label{conclusion}
In this article, we have proposed a DTE MADRL-based power control scheme to allocate transmit power in HetNets. 
In particular, we introduce a DTE scheme, which decides the transmit power for each AP simultaneously based on local observation.
In addition, only the interference power and achievable rates of neighbors are required in both training and testing phases, which is practical in real systems since the unavailable global and interference CSI is not required.
To promote cooperation among agents, we have proposed PQL algorithm for training MADRL systems, which has almost no extra cost compared with the simple and widely used IQL algorithm.
Simulation results show that with limited network layers and local observation, our proposed scheme is able to reach over 90\% performance of the conventional WMMSE algorithm and outperforms the existing DTE IQL and HQL algorithms under various system settings.


\end{document}